\documentclass[
twocolumn
,showpacs,preprintnumbers,amsmath,amssymb,prb,superscriptaddress]{revtex4}
\usepackage{amsmath}
\usepackage{dcolumn}% Align table columns on decimal point \usepackage{bm}%
\usepackage{bm} %bold math

\usepackage[dvipdfm]{graphicx}

\begin{document}

\title{Doping dependence of Hall coefficient and evolution of coherent electronic state
in the normal state of Fe-based superconductor Ba$_{1-x}$K$_{x}$Fe$_{2}$As$_{2}$}

\author{Kenya Ohgushi}
\affiliation{Institute for Solid State Physics, University of
Tokyo, Kashiwa, Chiba 277-8581, Japan} \affiliation{JST, TRIP,
Chiyoda, Tokyo 102-0075, Japan}

\author{Yoko Kiuchi}
\affiliation{Institute for Solid State Physics, University of
Tokyo, Kashiwa, Chiba 277-8581, Japan}

\date{\today}

\begin{abstract}
We investigated the in-plane transport properties of the Fe-based
superconductor Ba$_{1-x}$K$_{x}$Fe$_{2}$As$_{2}$ with a wide
composition range $0 \leq x \leq 0.55$. We show that the doping
dependence of the Hall coefficient is well-described by the
Boltzmann equation for a two-band system with a rigid-band
approximation. We successfully deduced transport parameters, which
suggested that holes with heavier mass conduct more smoothly
than electrons. Moreover, the temperature variation of the Hall
coefficient indicated that an anomalous coherent state
characterized by heavy quasiparticles in hole bands evolved below
$\sim100$ K, predominantly in the optimal and overdoped regions.
We argue that this phenomenon can be understood in relation to the
pseudopeak structure observed in angle-resolved photoemission
spectroscopy.

\end{abstract}

\maketitle

\section{INTRODUCTION}

In correlated electronic systems, we frequently observe a
crossover phenomenon which does not accompany the long-range order
associated with symmetry breaking. A famous example is the
pseudogap behavior in a normal state of cuprate superconductors,
where various quantities show a gap-like feature below a
characteristic temperature. \cite{Damascelli} Despite the
significant effort that has been made to date, there still remains
debate concerning the microscopic mechanism.

The pseudogap phenomena are also observed in a canonical system of
Fe-based superconductors Ba$_{1-x}$K$_{x}$Fe$_{2}$As$_{2}$,
\cite{rotter} an electronic phase diagram of which is shown in
Fig. 1. An angle-resolved photoemission spectroscopy (ARPES)
measurement for $x =0.25$ revealed a spectral weight transfer to a
deeper energy level of $\sim$18 meV in the innermost hole Fermi surface
below $\sim$ 120 K, indicating an opening of the psedogap. \cite{Xu}
On the other hand, a laser ARPES for
$x=0.41$ clarified the evolution of a pseudopeak structure
centered $\sim 12$ meV below the Fermi energy in the three hole
Fermi surfaces below $\sim 100$ K. \cite{Shimojima1} In this
course, the density of states at the Fermi energy gradually
increased upon cooling, which is in stark contrast to the pseudogap
phenomena. In order to unravel the underlying microscopic
mechanism of the pseudogap and pseudopeak features, we need to
collect detailed information on electronic states over a wide $x$
range.

A sensitive probe of a crossover phenomenon is the Hall
coefficient ($R_{\rm H}$), which can detect not only single
particle information but also the electron correlation effect.
Indeed, the temperature $(T)$ evolution of $R_{\rm H}$ was one of the
earliest pieces of evidence of a pseudogap in cuprate superconductors. \cite{Hwang} 
In Fe-based superconductors, however, the multiband
nature of Fermi surfaces, which consist of three hole bands and
two electron bands (see inset of Fig. 1), \cite{Yin} prevents us from
providing a straightforward interpretation of $R_{\rm H}$.
Moreover, most previous studies focused on $R_{\rm H}$ in alloyed
compounds such as Ba(Fe$_{1-y}$Co$_{y}$)$_{2}$As$_{2}$,
\cite{Rullier-Albenque1, Mun, Fang, Katayama, Rullier-Albenque2,
Olariu} where a disorder effect as well as the band
reconstruction were not negligible. Hence, a systematic study of
$R_{\rm H}$ for Ba$_{1-x}$K$_{x}$Fe$_{2}$As$_{2}$ has been highly
anticipated. Recently, such a study was actually performed; however, 
it still focused on the underdoped regime. \cite{Shen}   

\begin{figure}[t]
\includegraphics[scale=0.37]{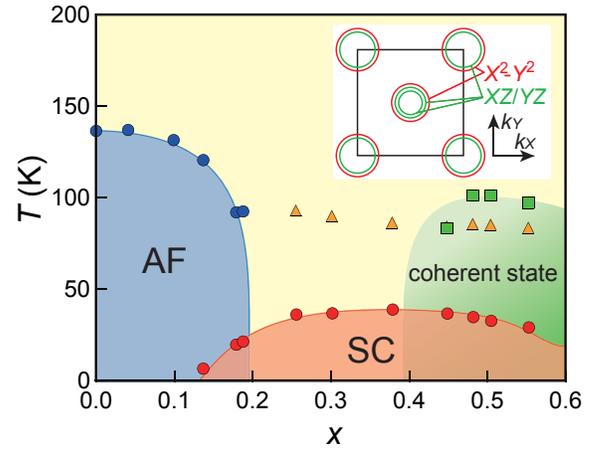}
\caption{(Color online) 
Electronic phase diagram of Ba$_{1-x}$K$_{x}$Fe$_{2}$As$_{2}$ in the 
K content ($x$) and temperature ($T$) plane, where AF and SC 
stand for antiferromagnetic and superconducting phases, respectively.
Triangles and squares indicate the crossover temperature obtained 
by the resistivity and Hall coefficient, respectively.
The inset is a schematic view of Fermi surfaces. The orbital character 
of each band is also shown. \cite{Yin}
}
\end{figure}

In this work, we investigated normal state of
Ba$_{1-x}$K$_{x}$Fe$_{2}$As$_{2}$ over a wide composition range 
$0 \leq x \leq 0.55$ by measuring in-plane transport properties 
with a special focus on $R_{\rm H}$. The analysis based
on the Boltzmann equation indicated that the quasiparticle mass
ratio of holes to electrons increased with decreasing $T$, which
likely cause a decrease in $R_{\rm H}$ below 
$\sim$100 K for $0.45 \leq x \leq 0.55$. We claim that this 
behavior is closely related to the evolution of the pseudopeak 
in the ARPES spectra, and argue that its mechanism can be 
understood in terms of a coherent quasiparticle formation 
due to the coupling with a boson in the hole bands.

\section{EXPERIMENT}

Single crystals of Ba$_{1-x}$K$_{x}$Fe$_{2}$As$_{2}$
were grown by the self-flux method as reported in the literature.
\cite{Suzuki} Since the composition of K has a tendency to be
spatially inhomogeneous even within a single crystal, we carefully
characterized each crystal. We mounted a crystal oriented along
the $c$ axis on an x-ray diffractometer and checked the 0 0 $l$
reflection width, which is a measure of $x$ variation according to
the linear relationship $c = 13.01 + 0.821 x$ \AA ($c$ being the
lattice parameter). The variation of $x$ in a crystal used for
transport measurements was less than $\pm 0.01$. The in-plane
resistivity ($\rho$) was measured by the standard four-probe
method. The Hall resistivity in the plane was measured by using a
rotator under a magnetic field of 7 tesla at a fixed value of temperature.

\section{RESULTS AND ANALYSIS}

\begin{figure}[t]
\includegraphics[scale=0.6]{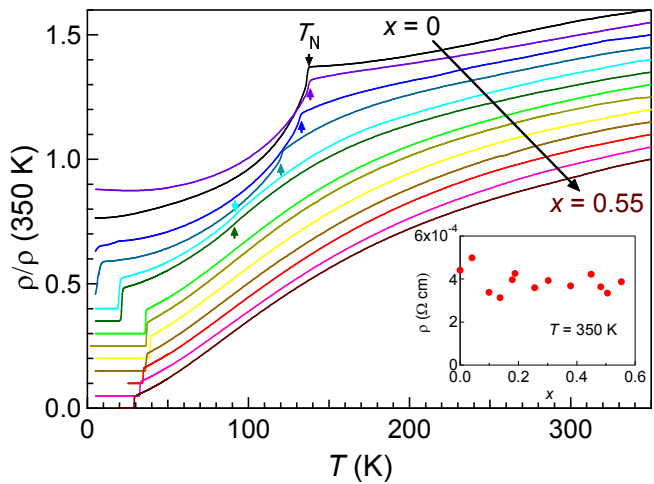}
\caption{(Color online) 
Temperature ($T$) dependence of resistivity ($\rho$) scaled 
by the 350 K value for Ba$_{1-x}$K$_{x}$Fe$_{2}$As$_{2}$. 
The composition is $x=0, 0.04, 0.10, 0.14, 0.18, 0.19, 0.26, 0.30, 
0.38, 0.45, 0.48, 0.51,$ and $0.55$ from the top to the bottom.
Note that each curve is shifted by 0.05 for clarity. 
The arrows indicate the antiferromagnetic transition temperature 
($T_{\rm N}$). The inset shows the $x$ dependence of $\rho$ at 350 K.  
}
\end{figure}

\subsection{Resistivity}

Figure 2 represents the $T$ variation of $\rho$ scaled by the 350
K value, which somewhat depends on $x$ (inset of Fig. 2). The quantity $\rho$
at $x= 0$ was weakly $T$ dependent in the paramagnetic state and
showed an abrupt decrease below 137 K, which corresponded to the
antiferromagnetic transition temperature $(T_{\rm N})$. Upon K
substitution, the anomaly appeared at a lower $T$ in a subtle
manner and completely disappeared at $x = 0.26$. In the wide
composition range $0.14 \leq x \leq 0.55$, the zero resistivity
was observed at low $T$. The superconducting transition
temperature $(T_{c})$ was estimated by the midpoint of the $\rho$
drop; a typical $T_{c}$ width was 0.6 K. We then obtained an
electronic phase diagram (Fig. 1) which was in perfect agreement with
the phase diagram for polycrystalline samples. \cite{Rotter} 
The quantity $\rho$ for $0.26 \leq x \leq 0.55$ exhibited a crossover 
from a high-$T$ upward-concave to a low-$T$ downward-concave 
behavior; the characteristic $T$ (80$-$95 K) are plotted as open 
triangles in Fig. 1.

The power-law behavior of $\rho$ is a key ingredient in relation to the possible 
quantum critical phenomena near an antiferromagnetic-paramagnetic 
phase boundary. We fit the data for 0.26 $ \leq x \leq$ 0.55, where 
the antiferromagnetic state disappears down to the lowest $T$, 
with the function $\rho = \rho_{0}+ A T^{n}$.
When we take a fitting range as $40-80$ K, this fitting yields 
$n =$ 1.62 for $x=$ 0.26 and $n =$ 1.47 for $x=$ 0.55. The deviation from the 
Fermi liquid value $n=2$ seemingly suggests an influence of 
critical antiferromagnetic fluctuations; however, we need special care. 
The $\rho_{0}$ value becomes negative at $x \geq$ 0.38, indicating that  
the actual $n$ is much larger than $n$ deduced by the fitting. One cannot 
make conclusions about the quantum critical behavior in the present system until
the normal-state $\rho$ is acquired down to lower $T$ by
eliminating the superconducting phase with a high magnetic field. We also point out the 
possibility that earlier studies reporting $n<2$ in Fe-based superconductors 
overlooked the large $T$ variation of $\rho$ at low-$T$ ranges owing to a 
large residual resistivity.

\subsection{Hall coefficient}

\begin{figure}[t]
\includegraphics[scale=0.70]{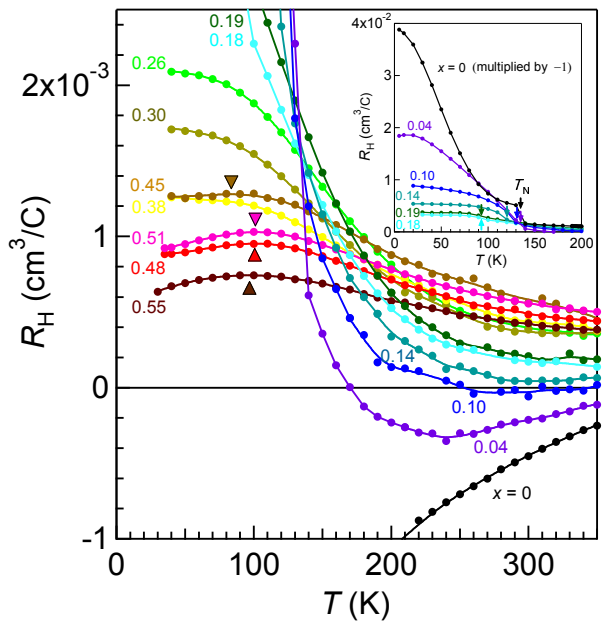}
\caption{(Color online) 
Temperature ($T$) dependence of Hall coefficient ($R_{\rm H}$) 
for Ba$_{1-x}$K$_{x}$Fe$_{2}$As$_{2}$ ($0 \leq x \leq 0.55$). 
The triangles indicate the crossover temperature, where 
$R_{\rm H}$ turns to a decrease on cooling. The inset shows $R_{\rm H}$ 
for samples which undergo the antiferromagnetic transition at low $T$ 
in a full scale. The arrows indicate the antiferromagnetic transition 
temperature ($T_{\rm N}$). Note that the data for $x=0$ are multiplied by $-1$.
}
\end{figure}

$R_{\rm H}$ displayed in Fig. 3 exhibits a significant variation
as a function of $x$ and $T$. At $x=0$, a negative $R_{\rm H}$
gradually increased in magnitude upon cooling, showed a
discontinuous jump at $T_{\rm N}$, and reached a constant value at
5 K. When holes are doped, $R_{\rm H}$ changes its sign. In a wide
$T$ range, $R_{\rm H}$ still showed a monotonic increase in
magnitude upon cooling; however, such a tendency became more
moderate with increasing $x$. For $ 0.45 \leq x \leq 0.55$, we
found a crossover phenomena from the high-$T$ increasing to the
low-$T$ decreasing behavior in $R_{\rm H}$ at $T^{*} \sim$100 K,
which is indicated by triangles in Fig. 3. The crossover
temperature $T^{*}$ is plotted as open squares in Fig. 1, where we
call the low-$T$ phase a coherent state. From now on, we
concentrate on $R_{\rm H}$ in the paramagnetic phase.

In order to interpret the complicated behavior of $R_{\rm H}$, we
first focus on the doping dependence. We introduce the effective
carrier density $n_{\rm eff}=\frac{V}{e} \frac{1}{R_{\rm H}}$,
where $V$ is the volume per Fe atom and $e$ is the elementary
electric charge, and plot them against $x$ in Fig. 4. This
quantity becomes $x/2$ if the parent compound ($x=0$) is a band
insulator. Actually, the low-$T$ data seemingly obeyed this
relationship (inset of Fig. 4); however, this was an
accidental agreement caused by the enhanced $R_{\rm H}$ in the
underdoped region. Instead, intrinsic information could be deduced
from the high-$T$ data, where $1/n_{\rm eff}$ increased from a
negative to a positive value with increasing $x$. We analyzed the 
350 K data by adopting the Boltzmann equation for a two-band system, \cite{multiband}
$\frac{1}{n_{\rm eff}}= \frac{n_{h} (\tau_{h}/m_{h})^2-n_{e}
(\tau_{e}/m_{e})^2}
{(n_{h}\tau_{h}/m_{h}+n_{e}\tau_{e}/m_{e})^2},$ where $n_{h}
(n_{e})$, $m_{h} (m_{e}) $, and $\tau_{h} (\tau_{e}) $ are the
carrier density per Fe atom, the quasiparticle mass, and the
relaxation time for the hole (electron) band, respectively.
\cite{Rullier-Albenque1, Fang, Katayama, Tsukada} 
In contrast to an alloyed system, a rigid band shift of the 
Fermi level from a compensated metal with $n_{h}=n_{e}=n_{0}$ at 
$x =0$ is reasonably anticipated in the current system.
Then, the carrier density changes in form in
which $n_{h} =n_{0}+\frac{x}{2} \frac{m_{h}}{m_{h}+m_{e}}$ and
$n_{e} =n_{0}-\frac{x}{2} \frac{m_{e}}{m_{h}+m_{e}}$. Substituting
these expressions into the $n_{\rm eff}$ formula, we obtained a
fitting function. We performed a fitting with the data at 350 K
under the assumption that the electron band vanishes at $x=1$ $[2
n_{0} (1+m_{h}/m_{e}) =1]$. \cite{constraint} The fitting quality
was fairly good, when $n_{0} = 0.12$, $m_{h}/m_{e}=3.1$, and
$\tau_{h}/\tau_{e}=2.6$ (Fig. 4).

The obtained transport parameters are compared with
the ARPES results. (1) The $n_{0}$ value corresponds to the sum of
hole or electron Fermi surface areas at $x=0$, which is reported
to be $0.06-0.10$. \cite{Brouet, Yi} This value is close to our
result, supporting the validity of the present analysis. (2) The
quasiparticle mass for each hole and electron band was determined
by ARPES measurements for $x=0.4$; \cite{Ding} a simple average
among the three hole and two electron bands leads to
$m_{h}/m_{e}=4.8$, which is consistent with our result. Although
the lighter mass of electrons than that of holes is considered to
primarily originate from a strong hybridization between Fe $3d$
and As $4p$ orbitals at the zone corner of the Brillouin zone,
\cite{Singh} we also speculate that an orbital-dependent correlation
effect selectively enhanced the mass of a particular band
and resulted in the mass asymmetry. \cite{Aichhorn} (3) The
relaxation time obtained by our analysis pointed to an apparently
peculiar conclusion: heavier holes experienced less dissipation
than electrons. The relevant scattering source is likely to be
phonons; \cite{scattering} the reason why the electron-phonon
interaction was so strong in electron bands is unclear at present.
A detailed analysis of the quasiparticle lifetime in the ARPES
spectra is therefore highly anticipated. (4) The mobility for hole
(electron) bands, which are defined as $\mu_{h}=\tau_{h}/m_{h}$
$(\mu_{e}=\tau_{e}/m_{e})$, had the ratio $\mu_{h}/\mu_{e}=0.84$;
$\mu_{h}<\mu_{e}$ was the reason for the negative value of $R_{\rm
H}$ at $x=0$.

We move to the $T$ dependence. We performed the
above-mentioned fitting with the data at $200 \leq T \leq 350$ K;
the results are summarized in Fig. 5. Upon
cooling, $n_{0}$ showed a decrease, whereas $m_{h}/m_{e}$ and
$\tau_{h}/\tau_{e}$ increased, keeping $\mu_{h}/\mu_{e}$ nearly
constant. We stress that $R_{\rm H}$ of underdoped samples was so
significantly enhanced in magnitude in the low-$T$ region that the
present analysis is accurate in the high-$T$ limit. Nevertheless,
the global $T$-dependence obtained might be meaningful and 
is expected to continue to $T<200$ K. Here, let us divide the $R_{\rm H}$
behavior with decreasing $T$ into two parts, (1) an increase in
magnitude observed in wide $x$ and $T$ regions and (2) a decrease
below $T^{*}$ observed in the specific composition $ 0.45 \leq
x \leq 0.55$, and identify the transport parameters responsible
for these behaviors. We claim that the former behavior was
predominantly induced by the decrease in $n_{0}$ and the latter
behavior was totally induced by the increase in $m_{h}/m_{e}$;
this was concluded by changing a transport parameter in the
Boltzmann equation, while other parameters remained fixed in a
relevant parameter region. Which of the two contradictory
behaviors is actually observed depends on a competition between
the two underlying mechanisms, which we will argue below.

\begin{figure}[t]
\includegraphics[scale=0.58]{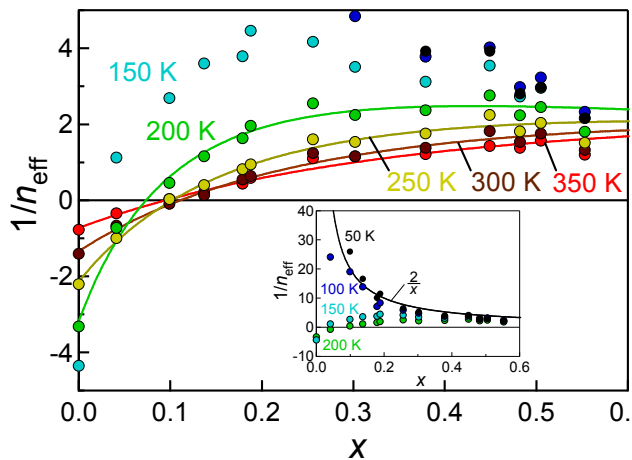}
\caption{(Color online) 
The doping dependence of the inverse effective carrier density 
($n_{\rm eff}$) for Ba$_{1-x}$K$_{x}$Fe$_{2}$As$_{2}$ ($0 \leq x \leq 0.55$). 
The high- and low-temperature ($T$) data are shown in the main 
panel and inset, respectively. The solid lines in the main panel 
represent results of fitting, whereas the solid line in the inset is 
a $n_{\rm eff} = x/2$ curve. 
}
\end{figure}

\section{DISCUSSION}

A gradual increase of $|R_{\rm H}|$ upon cooling is widely observed
near the antiferromagnetic phase in Fe-based superconductors, and
its origin is ascribed to antiferromagnetic fluctuations.
\cite{Fang, Katayama, Kasahara} The present data showing the
weakened $T$ variation of $R_{\rm H}$, with a departure from the
antiferromagnetic phase in the $x$-$T$ plane, supports this
conjecture. Our claim that a reduction in $n_{0}$ is responsible
for this behavior points to a shrinkage of the Fermi surface
areas; however, such a feature is indiscernible in the ARPES
spectra. \cite{Shimojima1} Instead, a more plausible scenario is
that the $n_{0}$ decrease is a pretension produced by a
phenomenological analysis; we need to directly deal with $R_{\rm
H}$ in terms of the pseudogap formation \cite{Xu} and/or the
vertex correction for the current, \cite{kontani} both of which
are supposed to originate from antiferromagnetic fluctuations.

The decrease of $|R_{\rm H}|$ on cooling is most likely relevant
to an evolution of the pseudopeak structure observed in the ARPES
spectra. \cite{Shimojima1}  There are two pieces of supporting evidence for this
interpretation. One is that the crossover $T$ of the two phenomena
is quite similar at $\sim$100 K. Another is that both phenomena 
are related to the increase of $m_{h}/m_{e}$: since the density of 
states is proportional to the quasiparticle mass in two dimensions,
the evolution of pseudopeak only in hole bands (not in electron bands) 
hints at the $m_{h}/m_{e}$ increase; \cite{Shimojima1} 
we also recall that our analysis connected the $|R_{\rm H}|$ decrease with the
$m_{h}/m_{e}$ increase. 
One might think that the composition at which the pseudopeak 
structure was observed $(x=0.41)$ is beyond $ 0.45 \leq x \leq 0.55$, 
where the $R_{\rm H}$ decrease was observed. However, 
the $R_{\rm H}$ increase upon cooling became moderate below $\sim 100$ K  
for $0.26 \leq x \leq 0.38$, hinting that the same phenomenon as observed 
in $ 0.45 \leq x \leq 0.55$ happens on the more underdoped side. 
We conclude that the coherent state is formed in a wider $x$ range, but 
is most stable in the optimal and overdoped regions. \cite{Malaeb} 

\begin{figure}[t]
\includegraphics[scale=0.55, clip]{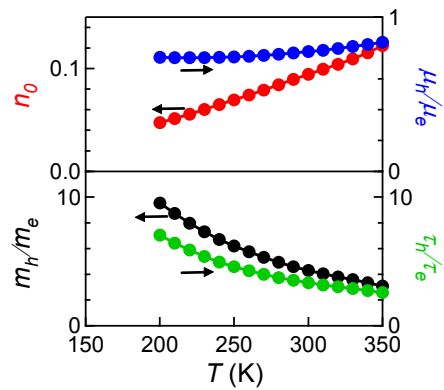}
\caption{(Color online) 
The temperature $(T)$ dependence of transport parameters 
obtained by fitting with the two-band Boltzmann equation 
for Ba$_{1-x}$K$_{x}$Fe$_{2}$As$_{2}$ ($0 \leq x \leq 0.55$). 
$n_{0}$ stands for the hole or electron density per Fe atom at $x=0$. 
$m_{h} (m_{e})$, $\tau_{h} (\tau_{e}) $, and $\mu_{h} (\mu_{e})$ 
are the quasiparticle mass, the relaxation time, and the mobility 
for the hole (electron) band, respectively.
}
\end{figure}

We now discuss the underlying microscopic mechanism of the increase 
of $m_{h}/m_{e}$, which consequently leads to the decrease of 
$|R_{\rm H}|$ below $T^{*}$. The simplest interpretation is that holes, 
coupling with a certain boson centered at $\sim$ 12 meV below 
the Fermi energy, formed heavy quasiparticles and
acquired quantum coherence below $T^{*}$. \cite{Shimojima1} A
question to be answered is what the boson is. 
A possible candidate is a magnon. However, the stable coherent
state separated from the antiferromagnetic phase (Fig. 1) does not
match this scenario. Thus, the nature of the coherent state below 
$T^{*}$, as well as its relevance to superconductivity, 
is unclear at present.

A comparison with other related systems merits a further
understanding. Whereas an increase of $|R_{\rm H}|$ upon cooling is
observed in various Fe-based superconductors, including
Ba(Fe$_{1-y}M_{y})_{2}$As$_{2}$ ($M=$ Co, Ni, Cu, and Ru)
\cite{Rullier-Albenque1, Mun, Fang, Katayama, Rullier-Albenque2,
Olariu} and P-substituted BaFe$_{2}$As$_{2}$, \cite{Kasahara} the
decrease of $|R_{\rm H}|$ is quite rare. An exceptional case is
$M=$ Co, which has a narrow composition range $0.09 \leq y \leq
0.15$, in which a negative $R_{\rm H}$ decreases at high $T$ and
tends to increase at $\sim 80$ K with decreasing $T$.
\cite{Rullier-Albenque1, Mun, Fang, Katayama} A close inspection 
of the electron-hole asymmetry will open
up a route to identify the boson relevant to the formation of the 
coherent electronic state.

Finally, we briefly comment on the relevance between the crossover
phenomenon associated with $\rho$ and $R_{\rm H}$ (open triangles
and open squares in Fig. 1, respectively). Although the two
characteristic $T$s are quite similar in the composition ranges
investigated, we concluded that they have distinguishable origins.
This is because of a lack of correspondence at $x=1.0$; that is,
$\rho$ exhibited a crossover at $\sim$70 K, whereas $R_{\rm H}$
monotonically increased on cooling (data not shown). The $T$
dependence of $\rho$ likely reflects the Bloch--Gr${\rm
\ddot{u}}$neisen formula due to the scattering by lattice
vibrations.

\section{CONCLUSIONS}
We investigated normal state properties of
Ba$_{1-x}$K$_{x}$Fe$_{2}$As$_{2}$ ($0 \leq x \leq 0.55$) by
systematic Hall coefficient measurements. The data analysis based
on the Boltzmann equation indicates that heavier holes conduct
smoothly while experiencing fewer dissipations than electrons.
We also revealed the formation of heavy coherent
quasiparticles in hole bands below $\sim 100$ K, predominantly in the
optimally-doped and overdoped regions.

\section*{ACKNOWLEDGMENT}
We thank T. Shimojima, W. Malaeb, K. Okazaki, K. Nakayama, T.
Sato, H. Fukazawa, and I. Tsukada for a fruitful discussion. This
work was supported by Special Coordination Funds for Promoting
Science and Technology, Promotion of Environmental Improvement for
Independence of Young Researchers.

%We should comment that our analysis totally rely on the two-band model, 
%which is incomplete to describe Fe-based superconductors with 
%five conduction bands. Our analysis becomes exact only when the relaxation 
%time and the band mass do not depend on the band index within 
%the hole (electron) bands. This situation is not realized in reality, 
%because there are several hole (electron) bands with distinguishable  
%orbital characters (inset of Fig. 1). Nevertheless, we believe in that 
%our conclusions concerning the asymmetry between electron and hole bands 
%are basically true, since $R_{\rm H}$ is quite sensitive to the carrier 
%types. 
%It is highly anticipated that high-resolution ARPES measurements 
%reveal transport parameters including the quasiparticle life time for 
%all the bands and deduce the conductivity tensor for each band. 


\begin{thebibliography}{}
\bibitem{multiband}
We adopted the two-band model, which is incomplete to describe Fe-based 
superconductors with five conduction bands. Our analysis becomes exact 
only when the relaxation time and the band mass do not depend on the band 
index within the hole (electron) bands. \cite{Katayama} 
This situation is not realized in reality, 
because there are several hole (electron) bands with distinguishable  
orbital characters (inset of Fig. 1). Nevertheless, we believe in that 
our conclusions about the asymmetry between hole and electron bands 
are basically true, since $R_{\rm H}$ is a sensitive probe concerning 
the carrier-type difference.
\bibitem{scattering}
We speculate that spin fluctuations play a minor role in $\rho$ at
350 K, since $\rho$ hardly vary for $0 \leq x \leq 0.55$, where
the antiferromagnetic fluctuation strength is expected to change
considerably (inset of Fig. 2).
\end{thebibliography}
\end{document}